\documentclass[pdflatex,sn-basic, Numbered,iicol]{sn-jnl}

\usepackage{graphicx}%
\usepackage{multirow}%
\usepackage{amsmath,amssymb,amsfonts}%
\usepackage{amsthm}%
\usepackage{mathrsfs}%
\usepackage[title]{appendix}%
\usepackage{xcolor}%
\usepackage{textcomp}%
\usepackage{manyfoot}%
\usepackage{booktabs}%
\usepackage{algorithm}%
\usepackage{algorithmicx}%
\usepackage{algpseudocode}%
\usepackage{listings}%
\usepackage{siunitx}%
\usepackage{subcaption}%
\usepackage{url}%
\usepackage{hyperref}%
\usepackage{array}%

\theoremstyle{thmstyleone}%

\theoremstyle{thmstyletwo}%

\theoremstyle{thmstylethree}%

\begin{document}

\title[Digitization Pipelines for Historiographical Sources]{A Comparative Evaluation of Digitization Pipelines for Historiographical Sources}

\author*[1]{\fnm{Marina} \sur{Gómez Rey}}\email{marinago@pa.uc3m.es}

\author*[1]{\fnm{Patricia} \sur{Callejo}}\email{pcallejo@it.uc3m.es}

\author[1]{\fnm{Mario} \sur{Muñoz-Organero}}\email{munozm@it.uc3m.es}

\author[1]{\fnm{Carlos} \sur{Alario-Hoyos}}\email{calario@it.uc3m.es}

\affil*[1]{\orgdiv{Telematics Engineering Department}, \orgname{Universidad Carlos III de Madrid}, \orgaddress{\city{Leganés}, \state{Madrid}, \postcode{28911}, \country{Spain}}}

\abstract{
\textbf{Purpose:} The digitization of historical documents presents fundamental challenges for modern information retrieval and Artificial Intelligence (AI) systems. Optical character recognition (OCR) errors in source corpora propagate through retrieval-augmented generation (RAG) pipelines, compromising the factual accuracy of generated outputs.

\textbf{Methods:} This study presents a systematic evaluation of PDF-to-text extraction pipelines applied to historiographical secondary sources on the Visigothic period. We assess thirteen distinct approaches spanning three methodological families: direct extraction, Large Language Model (LLM) post-correction, and chunk-and-extract. Documents are stratified into five categories based on production method and visual complexity. Performance is measured using character error rate (CER) and word error rate (WER) against manually corrected ground truth.

\textbf{Results:} Results demonstrate that direct extraction with Marker achieves superior performance (98.70\% CER accuracy; 97.71\% WER accuracy overall), while conventional OCR pipelines exhibit substantial degradation on scanned documents and complex layouts. Embedded-text extraction performs well on digital PDFs but fails on scanned documents. LLM post-correction does not provide systematic improvements and frequently degrades accurate extractions.

\textbf{Conclusion:} End-to-end document parsing is the most reliable approach for heterogeneous historical collections. Document characteristics such as scan quality, layout complexity, and the presence of embedded text layers have a significant impact on extraction accuracy. LLM-based post-correction should not be assumed beneficial by default and requires validation before large-scale application.
}


\maketitle

\section{Introduction}\label{sec1}

Information quality fundamentally determines the reliability of knowledge systems \cite{Berger2020}. In text-based applications, degradation introduced at the digitization stage can cascade through subsequent processing layers, affecting search, retrieval, and knowledge generation tasks. This dependency becomes critical when historical documents serve as knowledge sources for modern Artificial Intelligence (AI) systems, especially Retrieval-Augmented Generation (RAG) architectures that enhance Large Language Model (LLM) outputs with retrieved contextual information \cite{lewis2020retrieval, Livathinos2025}.

RAG-based systems operate through a multi-stage process: documents are first segmented into smaller chunks, which are then transformed into vector representations using embedding models, and subsequently stored in vector databases. When a user submits a query, the system retrieves semantically relevant chunks and supplies them as context to an LLM, which generates responses conditioned on both the query and retrieved information. This architecture has become foundational for conversational AI systems, enabling question answering over specific document collections, supporting domain-specific applications, and grounding model outputs in external knowledge sources rather than relying solely on pretrained parameters.

However, the effectiveness of RAG-based systems depends critically on the quality of the underlying document corpus. When source documents contain OCR character substitutions, word segmentation failures, or layout corruption, these errors propagate through the retrieval and generation stages \cite{zhang2025ocr}. RAG systems may fail to match queries to relevant chunks due to corrupted entity names or key terms. Even when retrieval succeeds, corrupted content is passed directly to the generation model, which may reproduce or amplify errors in its outputs. Because modern LLMs typically generate fluent responses independent of source quality, users may be unaware that responses are grounded in corrupted content.

Historical texts present distinct challenges for OCR systems integrated into RAG-based systems. Unlike contemporary digital documents, historical materials frequently exhibit multi-column layouts, non-standard typographical conventions, archaic orthography, physical degradation, and mixed content types \cite{Khan2024}. These characteristics introduce systematic errors including character substitutions, word segmentation failures, reading order corruption, and layout misinterpretation. When such errors persist in digitized corpora, they propagate through information retrieval pipelines and may be reproduced by LLMs, creating factually incorrect outputs that appear authoritative to end users.

Consider a representative failure case: when queried ``Who was deposed by Witérico?'', a RAG-based AI system responds ``Liuva 11'' rather than the correct ``Liuva II''. This error originates from OCR misrecognition of the Roman numeral II as the Arabic number 11. While superficially minor, such errors systematically corrupt historical knowledge in systems deployed to millions of users. The error becomes particularly relevant because modern conversational AI systems present outputs with high confidence regardless of source quality, rendering digitization errors invisible to non-expert users.

Despite advances in neural OCR architectures and vision-language models \cite{Fleischhacker2025}, no existing solution reliably digitizes complex historiographical documents. Recent work has explored LLM-based post-correction \cite{Kanerva2025, Sastre2025, Levchenko2025}, hybrid layout analysis approaches \cite{Fleischhacker2025}, and end-to-end document parsing \cite{Livathinos2025}. However, systematic comparative evaluations on heterogeneous historical corpora remain limited. This study addresses three research questions:

\textbf{RQ1:} How do contemporary PDF-to-text extraction pipelines perform across different document types in historiographical secondary sources, and what error patterns characterize each approach?

\textbf{RQ2:} To what extent does document heterogeneity (production method, visual complexity, scan quality) affect extraction accuracy across pipeline architectures?

\textbf{RQ3:} To what extent does LLM-based post-correction improve OCR quality for historical documents?

We present a systematic evaluation of thirteen extraction pipelines on a stratified corpus of historiographical secondary sources on the Visigothic period. This corpus is selected because it exhibits the heterogeneity typical of digital humanities collections, including poor quality scans, modern digitizations, native digital PDFs, and complex layouts, with sufficient linguistic complexity (Latin terms, proper nouns, Roman numerals) to stress-test OCR systems beyond standard benchmarks. We measure character and word level accuracy against manually corrected ground truth, with evaluation focused on the quality requirements of modern AI systems, particularly RAG architectures where OCR errors in source documents propagate directly to generated outputs. This work makes three key contributions. First, we provide empirical performance comparison of thirteen extraction pipelines across five document categories, revealing substantial variation in accuracy and characteristic failure modes. Second, we demonstrate how document characteristics (production method, scan quality, and layout complexity) systematically affect extraction quality, with direct implications for workflow design and tool selection. Third, we present evidence that LLM-based post-correction, despite theoretical promise, can degrade performance under standard edit distance metrics.

\section{Related Work}\label{sec2}

OCR technology has evolved from rule-based pattern matching systems to neural architectures trained on large document corpora. Contemporary approaches leverage convolutional neural networks for character recognition combined with recurrent architectures for sequence modeling, substantially improving accuracy over traditional methods. However, performance varies significantly across document types and quality levels. Tesseract, a widely deployed open-source engine, uses LSTM-based recognition following image preprocessing and layout analysis, with a Connectionist Temporal Classification (CTC) loss function \cite{Smith2013}. EasyOCR represents an alternative approach, employing end-to-end convolutional architectures that bypass explicit segmentation stages~\cite{jaidedai2022easyocr}. The effectiveness of these systems on historical documents with complex layouts and degraded quality remains an open question, particularly relevant in recent RAG-based systems where extraction errors can corrupt retrieved context and generated outputs.

Historical documents pose distinct challenges for OCR systems. Khan et al.~\cite{Khan2024} survey AI approaches for historical document transcription, identifying three challenges: non-standard typography, physical degradation, and archaic linguistic forms. Martinek et al.~\cite{Martinek2020} address data scarcity by combining fully convolutional segmentation networks with recurrent recognition models, achieving competitive accuracy on 19th-century German Fraktur texts with minimal training data. Fleischhacker et al.~\cite{Fleischhacker2025} show that explicit layout analysis prior to OCR substantially improves extraction quality for multi-column 19th-century documents, demonstrating that structure detection before text recognition outperforms direct OCR approaches on complex layouts. Accurate extraction of reading order requires thorough analysis of document layout. Breuel~\cite{Breuel2003} introduced whitespace rectangle-based approaches for document structure detection. Contemporary systems have evolved toward unified architectures that jointly perform layout detection, table recognition, and reading order determination~\cite{zhang2024document}.

Large language models have been proposed for OCR error correction. Kanerva et al.~\cite{Kanerva2025} evaluate LLM-based post-correction on historical Finnish and Swedish documents, finding that improvements depend on base OCR quality and language characteristics. Sastre et al.~\cite{Sastre2025} show that constraining LLM outputs to valid lexical items reduces hallucination while preserving correction capability. Levchenko~\cite{Levchenko2025} proposes frameworks for evaluating LLM performance on historical OCR, emphasizing domain-appropriate metrics. These studies reveal tensions between linguistic plausibility and preservation of original text.

Document quality fundamentally determines RAG system reliability, as retrieval quality directly determines generation quality. Zhang et al.~\cite{zhang2025ocr} demonstrate that OCR errors cascade through RAG pipelines, corrupting both retrieval and generation stages. Liu et al.~\cite{liu2024lost} demonstrate that document positioning within prompts significantly affects LLM attention distribution and accuracy, while Cuconasu et al.~\cite{cuconasu2024power} show that both document type and retrieval position influence RAG effectiveness. As AI systems scale to incorporate digitized archives and scholarly repositories, OCR quality emerges as a critical challenge for factual accuracy and reliability. Our work provides systematic evaluation of PDF extraction pipelines on historiographical documents, measuring character and word level accuracy to inform tool selection for digitization workflows that feed RAG systems.

\section{Methodology}\label{sec3}

This section describes the experimental methodology for evaluating PDF-to-text extraction pipelines on historiographical sources. We first detail corpus construction and document categorization, explain ground truth preparation, describe each evaluated pipeline with its configuration, provide tool selection rationale, and finally present the evaluation metrics.

\subsection{Corpus Construction and Selection}\label{subsec1}

We curated a corpus of historiographical secondary sources on the Visigothic kingdom (415--721 CE) from Spanish academic publications. This corpus was selected for three reasons. First, Spanish academic publications from the late 20th and early 21st centuries exhibit the heterogeneity typical of digital humanities collections: poor quality scans, modern digitizations, native digital PDFs, and complex multi-column layouts. Second, the corpus contains sufficient linguistic and typographical complexity to stress-test OCR systems: Latin terminology, proper nouns, Roman numerals, specialized historical vocabulary, and non-standard notation such as dagger symbols and superscript footnotes. Third, historiographical texts present a critical use case for digitization quality, as minor OCR errors (misrecognized dates, corrupted proper nouns, confused Roman numerals) can introduce significant factual inaccuracies when these documents feed information retrieval or RAG systems.

The selected corpus comprises 14 documents~\cite{orlandis1962poder,peman1950mtextordfeminine,orlandis1989cronica,guzman1989recaredo,lopez1987hispania,davila1991clases,salinero2023recaredo,moreno1993capitulos,cabrerizo2019dialogo,arias2020hacia,orlandis2000doble,masana1991hispania,rovira1992semblanzas,moreno2014organizacion}, yielding 15 document units since one source is split into two fragments. These are distributed across five categories of three documents each. This selection ensures that the evaluation captures the range of challenges that practitioners encounter when digitizing historical scholarship.

\begin{figure}[t]
  \centering
  \includegraphics[width=\columnwidth]{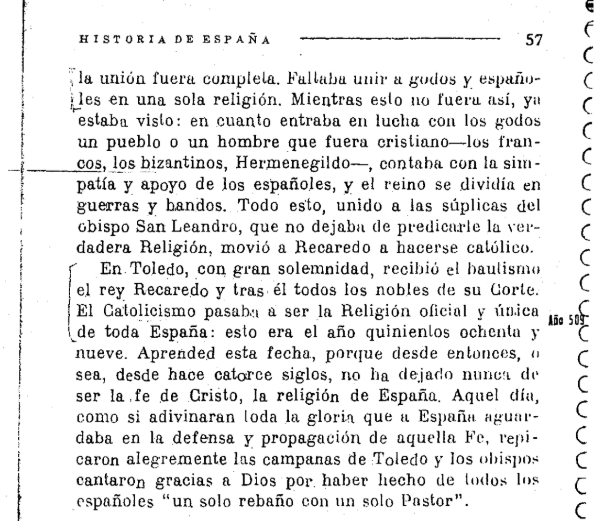}
  \caption{Example of Type 1 document.}
  \label{fig:type1}
\end{figure}

\textbf{Type 1: Low-Quality Scans with Annotations:} These documents are degraded scans from older print publications. They contain manual marginalia, underlining, physical stains, and other artifacts introduced during the scanning process or present in the original printed copies. The image quality is generally poor, with uneven lighting, skewed pages, and low contrast between text and background, which are common errors in historical texts \cite{sulaiman2019degraded}. These represent worst-case scenarios for automated digitization and are included to test the robustness of each pipeline under adverse conditions. An example is shown in Figure~\ref{fig:type1}.

\begin{figure}[t]
  \centering
  \includegraphics[width=\columnwidth]{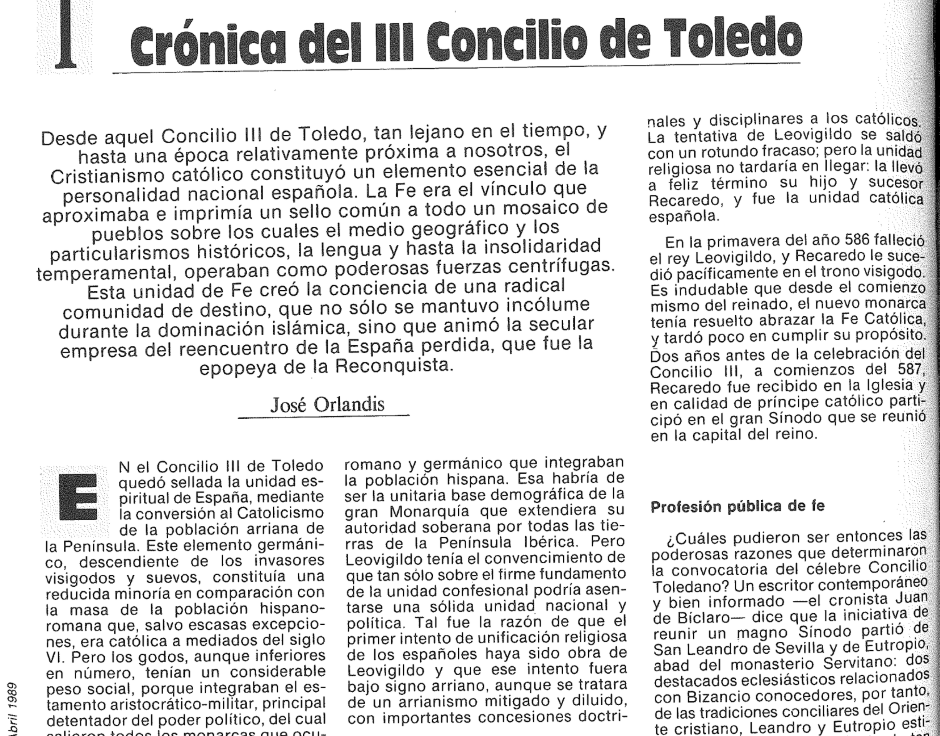}
  \caption{Example of Type 2 document.}
  \label{fig:type2}
\end{figure}

\textbf{Type 2: Multi-Column Complex Layouts:} These documents come from magazines, journals, or newspapers and feature two-column or multi-column layouts, embedded figures, captions, and non-linear reading order. They are included specifically to test the layout analysis capabilities of each pipeline, as incorrect column detection leads to reading order corruption where text from adjacent columns is interleaved. An example is shown in Figure~\ref{fig:type2}.

\begin{figure}[t]
  \centering
  \includegraphics[width=\columnwidth]{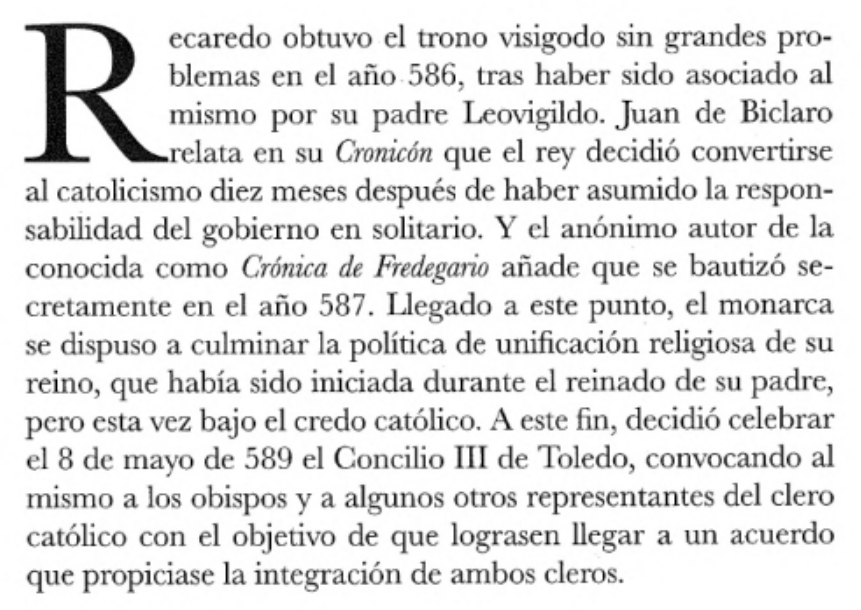}
  \caption{Example of Type 3 document.}
  \label{fig:type3}
\end{figure}

\textbf{Type 3: Clean Scans:} These documents are high-quality scans from modern publications with minimal visual artifacts and standard single-column layouts. Although they originate from scanned print sources, the scanning quality is good, with clear typography, high contrast, and minimal noise. These documents establish baseline performance on favorable historical materials, representing the best-case scenario for scanned documents. An example is shown in Figure~\ref{fig:type3}.

\begin{figure}[t]
  \centering
  \includegraphics[width=\columnwidth]{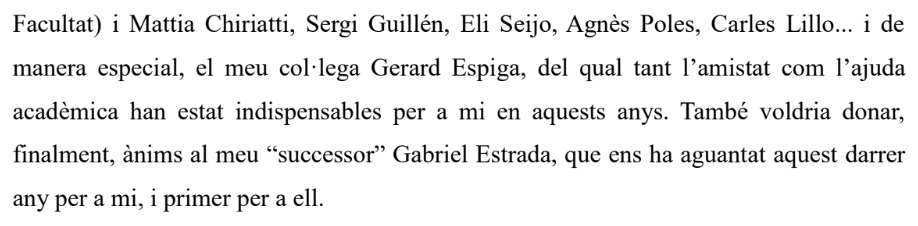}
  \caption{Example of Type 4 document.}
  \label{fig:type4}
\end{figure}

\textbf{Type 4: Digital PDFs:} These documents were originally created digitally (e.g., in Microsoft Word) and exported to PDF with embedded text layers. Because the text was never printed and scanned, these documents contain no visual degradation and their embedded text layers should be a faithful representation of the original content. They are included to test the performance of embedded-text extraction approaches under ideal conditions and to provide an upper bound for extraction quality. An example is shown in Figure~\ref{fig:type4}.

\begin{figure}[t]
  \centering
  \includegraphics[width=0.9\columnwidth]{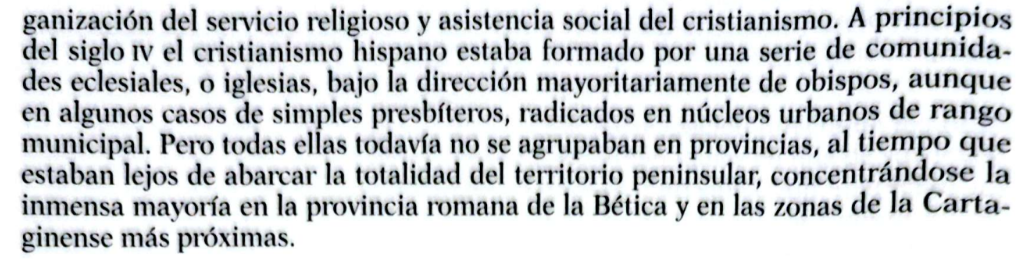}
  \caption{Example of Type 5 document.}
  \label{fig:type5}
\end{figure}

\textbf{Type 5: No Embedded Text:} These documents are scanned documents that, by their nature, do not contain embedded text layers. Unlike previous document types, where some scanned PDFs may include an OCR-generated text layer added during the scanning process, these documents provide only image data. Any pipeline that relies solely on extracting embedded text will produce empty output for these documents, making them a critical test for distinguishing between true OCR capability and mere text layer extraction. An example is shown in Figure~\ref{fig:type5}.

\subsection{Ground Truth Construction}\label{subsec2}

The ground truth was constructed manually by transcribing all text contained in the 15 selected documents. To manage workload while maintaining representativeness, we selected approximately 15 pages from each document. These manual transcriptions ensure maximum accuracy and avoid errors from automated processing. The resulting manually edited transcriptions were saved as plain text files for subsequent use in the evaluation of the pipeline outputs.

\subsection{Evaluated Pipelines}\label{subsec3}

We evaluated thirteen PDF-to-text extraction pipelines organized into three methodological families. The evaluation follows a comparative design where the thirteen extraction pipelines process the same corpus and their outputs are compared against the manually corrected ground truth. We organize the pipelines into three methodological families based on their processing approach: \textit{direct extraction} applies text extraction or OCR without post-processing; \textit{LLM post-correction} adds language model correction to direct extraction outputs; \textit{chunk-and-extract} segments pages into visual blocks before applying OCR or vision models to each block. This organization isolates the contribution of each processing component (text extraction, layout analysis, OCR engine, post-correction) and reveals how these interact with document characteristics. We focus exclusively on open-source tools for reproducibility and to reflect the practical constraints of digital humanities projects, where commercial API costs may be prohibitive for large-scale digitization. Below, we describe each pipeline and its configuration in detail.

\subsubsection{Direct Extraction}\label{subsubsec1}

The first approach relied on using extraction tools directly, without performing any subsequent post-processing actions. Three main tools were used: PyMuPDF \cite{artifex2024pymupdf}, Docling \cite{ibm2024docling}, and Marker \cite{datalab2024marker}, with Docling evaluated under three different configurations.

\textbf{PyMuPDF} is a Python library for extracting, converting, and manipulating PDF documents. It provides a set of functionalities including parsing PDF and other document formats, extracting text and images from pages, modifying document content by inserting, deleting, or rearranging pages and annotations, and retrieving text in different output formats. PyMuPDF also supports advanced document analysis features such as handling annotations and form fields, converting documents between formats, and optional OCR support. Additionally, it includes integration capabilities for pipelines such as RAG/LLM workflows.

For this study, the text extraction was performed using the \texttt{to\_markdown} function from the \texttt{pymupdf4llm} library, which converts PDF content to Markdown format. This approach is fast and straightforward but completely dependent on embedded text existence and quality, failing on scanned documents without text layers.

\textbf{Docling} is an open-source document processing and parsing toolkit originally developed by IBM, designed to integrate with the generative AI ecosystem \cite{Livathinos2025}. Its main objective is to convert a wide variety of document formats into structured, machine-readable representations that can be consumed by Generative AI systems. It supports end-to-end document workflows including RAG, question answering, and LLM-based analysis.

Docling provides parsing of a wide range of formats with advanced capabilities for layout understanding, table structure recognition, and OCR for scanned content. It offers integration with AI frameworks and provides utilities for document chunking and serialization into different output formats including Markdown, HTML, and JSON. We evaluated three configurations:

\begin{enumerate}\itemsep0em
    \item \textbf{Docling (No OCR):} Configured with \texttt{do\_ocr=False}, extracts text from document structure while preserving formatting through Markdown symbols. Behaves similarly to PyMuPDF but maintains structural information.
    \item \textbf{Docling + Tesseract:} Configured with \texttt{do\_ocr=True} using Tesseract~\cite{Smith2013}, which applies LSTM-based recognition with preprocessing, layout analysis, and region segmentation. Performance typically degrades on noisy images, complex layouts, or unusual fonts.
    \item \textbf{Docling + EasyOCR:} Configured with \texttt{do\_ocr=True} using EasyOCR, a convolutional neural network architecture that bypasses explicit segmentation. Expected to handle noisy images better but weaker at layout analysis for complex page structures.
\end{enumerate}

\textbf{Marker} is a document parser developed by Datalab that converts PDFs, images, and office documents into structured formats such as Markdown, JSON, or HTML. Unlike traditional pipelines that chain separate extraction and processing tools, Marker takes an end-to-end approach built on the Surya model suite, a set of vision transformers trained specifically for document understanding.

For layout detection, Marker uses a modified version of EfficientViT, a lightweight vision transformer architecture; for text recognition, it adapts Donut (Document Understanding Transformer)~\cite{kim2022donut}, a model that reads text directly from document images without a separate OCR step. Both models were trained from scratch to handle complex elements such as tables, equations, and multi-column layouts in over 90 languages.

Internally, Marker follows a three-stage pipeline. First, specialized models extract raw content, detecting layout regions, recognizing text, parsing table structures, and identifying equations. Second, a set of processors refine this output by merging fragmented text spans, correcting hyphenation artifacts, and establishing reading order. Third, renderers convert the processed content into the desired output format.

A key configuration option for our purposes is forcing OCR on every page, even when the PDF already contains embedded text. This ensures consistent extraction quality across both digital and scanned documents.

\subsubsection{LLM Post-Correction}\label{subsubsec2}

LLM post-correction pipelines add language model error correction to direct extraction outputs. The motivation is that LLMs can identify and reduce common OCR errors such as misspellings, incorrect word segmentation, and misplaced punctuation through their understanding of language patterns.

In this workflow, the text obtained from the previous extraction stage is used as input to an LLM, which is prompted explicitly to correct the provided text and return a cleaned version. The model selected for this task was Qwen3 (8B parameters) \cite{qwen2025}, chosen because it provides a good balance between performance and computational resource requirements, delivering acceptable results without the need for large-scale infrastructure.

The pipeline operates as follows: the raw OCR text is first segmented into paragraphs. Each paragraph is then processed independently by the LLM, which receives a carefully engineered prompt instructing it to correct errors and return cleaned text. Processing paragraphs independently helps manage context lengths and allows the model to focus on localized corrections, reducing the risk of hallucinations and improving overall results.

Creating an effective prompt is essential in Generative AI tasks, and following prompt engineering guidelines can significantly improve task outcomes. Common best practices include providing a clear set of instructions, properly separating sections within the prompt, using techniques such as few-shot prompting, and specifying explicit constraints on the output format. The prompt used for this task is the following:

\begin{verbatim}
You are an expert linguist in Spanish 
and your task is to **correct text 
coming from OCR**.

**RULES**:
- Keep the original content exactly as 
  it is, without inventing any 
  information.
- Correct incorrect spacing, misplaced
  hyphens, and broken or joined words.
- Fix accents and typographical errors.
- Correct words with spelling mistakes.
- Do NOT modify proper names except to
  correct accents.
- Do not rewrite or summarize: only 
  correct.

**Text to correct**:
"""
{paragraph}
"""

Return **only the corrected text**, 
with no comments, explanations, 
additional text, or extra words. 
Do not include "Corrected:" at the 
beginning or add any extra text at 
the end.
\end{verbatim}

Five post-correction pipelines were evaluated, each combining a different direct extraction method with the same LLM correction step.
\begin{enumerate}
    \item \textbf{PyMuPDF + LLM}: PyMuPDF extraction followed by paragraph-level LLM correction.
    \item \textbf{Docling (No OCR) + LLM}: Docling embedded-text extraction followed by LLM post-processing.
    \item \textbf{Docling + Tesseract + LLM}: Docling with Tesseract OCR extraction followed by LLM correction.
    \item \textbf{Docling + EasyOCR + LLM}: Docling with EasyOCR extraction followed by LLM correction.
    \item \textbf{Marker + LLM}: Marker extraction followed by LLM correction.
\end{enumerate}

\subsubsection{Chunk and Extract}\label{subsubsec3}

Inspired by Fleischhacker et al.~\cite{Fleischhacker2025}, who demonstrate that detecting document layout before text extraction significantly improves results, a two-step process was implemented: first, the document was segmented into chunks, and second, the text was extracted from each chunk.

The chunking procedure converts each PDF page into an image using PyMuPDF and OpenCV, an open-source computer vision library that provides the image processing primitives used throughout the pipeline. The DPI (dots per inch) used for this conversion is a critical parameter: values that are too low reduce image quality and produce fewer chunks, while values that are too high cause excessive fragmentation. We used values between 350 and 450 DPI depending on document complexity. The page image then undergoes contrast enhancement and morphological operations that merge nearby characters into coherent text regions. Contours are detected around these regions and extracted as individual image files, each representing a text block.

In practice, some blocks still contained too many elements, for instance, an entire page extracted as a single block. To address this, the chunking function was made recursive. Each block is evaluated against three criteria: (1)~aspect ratio exceeding 2.5, suggesting multiple columns; (2)~ink density below 40\%, measured through the Otsu thresholding method~\cite{otsu1975threshold}, an automatic binarisation technique that separates foreground text from background, indicating whitespace between distinct regions; and (3)~more than three disconnected contours, indicating multiple independent text sections. If any criterion is met, the block is re-segmented.

The recursive process introduced duplicate blocks, which are removed. Some visual noise present in the original scanned pages is also extracted as blocks, which must be handled during text extraction.

With the blocks prepared, the next step is text extraction. Three approaches were evaluated.

\begin{enumerate}
    \item \textbf{Chunk + MiniCPM-V:} Each extracted block is submitted to the MiniCPM-V~\cite{openbmb2024minicpm} vision model (8B parameters). Vision models are designed to process images as input accompanied by a text prompt specifying the desired output. The prompt used instructs the model to: (1)~first verify whether the image contains text, if not, return ``NO TEXT'', and (2)~if text is present, return only the original text, keeping paragraphs separated, without inventing words, completing cut-off sentences, or adding any text not present in the image. The ``NO TEXT'' instruction accounts for the fact that some blocks consist solely of noise from the original documents.
    \item \textbf{Chunk + Qwen3-VL:} The same chunking and prompting strategy is applied using the Qwen3-VL~\cite{qwen2025} vision model (4B parameters). This smaller model was included to evaluate whether a more lightweight vision model could achieve comparable extraction quality.
    \item \textbf{Chunk + EasyOCR:} Instead of a vision model, direct OCR is applied to each block using EasyOCR. Since the main limitation of EasyOCR is layout detection, while its strength lies in accurate text extraction at the word level, and the layout has already been determined through the chunking process, EasyOCR should produce more reliable word-level extraction when applied to individual, well-segmented blocks.
\end{enumerate}

\subsection{Tool Selection Rationale}\label{subsec4}

The tools evaluated in this study were selected based on their importance in the open-source document parsing ecosystem, as evidenced by community adoption and their established role as baselines in the Document AI and Vision-Language Model (VLM) literature.

For OCR engines, we include two complementary approaches. Tesseract~\cite{google2024tesseract, smith2007}, with over 72,000 GitHub stars, is the most widely deployed open-source OCR engine and serves as the fundamental baseline for document digitization. It relies on a Long Short-Term Memory (LSTM) architecture and represents the traditional approach to text recognition. EasyOCR~\cite{jaidedai2022easyocr, baek2019character}, with approximately 29,000 stars, represents the deep learning approach to scene text recognition. It combines a CRAFT detector with a ResNet-LSTM-CTC classifier, offering superior performance on non-standard fonts compared to traditional engines.

For PDF parsing, we evaluate three tools spanning different design philosophies. PyMuPDF~\cite{artifex2024pymupdf, artifex2024docs} dominates Python PDF text extraction with over 41 million monthly PyPI downloads~\cite{pypistats_pymupdf}. Rather than performing OCR, it accesses the underlying PDF object tree directly, enabling high-fidelity extraction of text and vector graphics from digitally born documents. Docling~\cite{ibm2024docling, auer2024doclingtechnicalreport}, which has accumulated over 52,000 GitHub stars since its release in 2024, was selected for its sophisticated layout analysis and table recognition capabilities, as well as its direct integration with multiple OCR engines. Marker~\cite{datalab2024marker}, with approximately 32,000 stars, employs the Surya engine for unified layout analysis and text recognition, including specialized handling of mathematical formulas through LaTeX reconstruction.

For vision-language models, we selected MiniCPM-V~\cite{yao2024minicpmvgpt4vlevelmllm}, which has demonstrated performance equivalent to commercial models such as GPT-4V. For LLM post-correction, we chose Qwen3-8B~\cite{qwen2025} because it achieves performance comparable to models with nearly twice its parameter count, specifically matching Qwen2.5-14B across several benchmarks, while remaining deployable on consumer hardware through Ollama and released under the Apache 2.0 license, ensuring full reproducibility.

\subsection{Evaluation Metrics}\label{subsec5}

We employ edit distance-based metrics standard in OCR evaluation literature:

\noindent\textbf{Character Error Rate (CER):} Levenshtein distance at the character level, computed as $\text{CER} = \frac{S_c + D_c + I_c}{N_c} \times 100$, where $S_c$, $D_c$, $I_c$ denote character-level substitutions, deletions, and insertions, and $N_c$ is the total number of characters in the ground truth.

\noindent\textbf{Word Error Rate (WER):} Levenshtein distance at the word level, computed as $\text{WER} = \frac{S_w + D_w + I_w}{N_w} \times 100$, where $S_w$, $D_w$, $I_w$ denote word-level substitutions, deletions, and insertions, and $N_w$ is the total number of words in the ground truth.

\noindent\textbf{CER Accuracy:} Reported as $1 - \text{CER}$, representing the fraction of correctly recognized characters. Negative values occur when insertions exceed ground truth length.

\noindent\textbf{WER Accuracy:} Reported as $1 - \text{WER}$, representing the fraction of correctly recognized words.

Following \citet{Levenshtein1966} and \citet{Berger2020}, we use standard dynamic programming implementations for edit distance computation. Macro-averages are computed across documents within each type and overall across the full corpus. This approach weights each document equally regardless of length, providing balanced performance assessment across heterogeneous materials.

\section{Results}\label{sec4}

\subsection{Overall Performance}\label{subsec6}

Table~\ref{tab:overall} presents macro-averaged accuracy across all document types. Marker achieves the highest overall performance (98.70\% CER accuracy; 97.71\% WER accuracy), substantially outperforming all alternatives. Among conventional OCR approaches, Docling + Tesseract yields the strongest results (89.06\% CER accuracy; 82.28\% WER accuracy), followed by Docling + EasyOCR (85.39\% CER; 78.07\% WER).

\begin{table}[]
\caption{CER and WER accuracy (\%) across all document types.}
\label{tab:overall}
\begin{tabular*}{\columnwidth}{@{\extracolsep{\fill}}lcc@{}}
\toprule
\textbf{Pipeline}            & \textbf{CER}           & \textbf{WER}  \\ \midrule
PyMuPDF*                    & 84.05                  & 79.43                  \\
Docling (No OCR)*           & 83.42                  & 77.85                  \\
Docling + EasyOCR           & 85.39                  & 78.07                  \\
Docling + Tesseract         & 89.06                  & 82.28                  \\
Marker                      & 98.70                  & 97.71         \\
PyMuPDF + LLM               & 55.95                  & 53.82                  \\
Docling (No OCR) + LLM      & 77.08                  & 73.29                  \\
Docling + EasyOCR + LLM     & 74.58                  & 67.40                  \\
Docling + Tesseract + LLM   & 78.77                  & 73.91                  \\
Marker + LLM                & 82.82                  & 79.30                  \\
Chunk + MiniCPM-V           & 9.36                   & 2.02                   \\
Chunk + Qwen3-VL            & 47.76                  & 43.72                  \\
Chunk + EasyOCR             & 52.08                  & 42.72                  \\ \bottomrule
\end{tabular*}
\par\vspace{2pt}
\parbox{\columnwidth}{\scriptsize *These pipelines rely on embedded text and yield 0\% accuracy on Type~5 documents; these values are excluded from the average.}
\end{table}

Embedded-text extraction approaches (PyMuPDF, Docling No OCR) show moderate overall performance but completely fail on documents without embedded text layers (Type 5), yielding 0\% accuracy. Their overall scores are therefore artifacts of corpus composition rather than true capability measures. When considering only documents with embedded text (Types 1--4), PyMuPDF achieves 84.05\% CER accuracy and 79.43\% WER accuracy, while Docling (No OCR) achieves 83.42\% CER and 77.85\% WER.

LLM post-correction pipelines exhibit mixed results. All LLM-augmented pipelines show degraded overall performance relative to their base OCR systems. For example, Docling + Tesseract achieves 89.06\% CER accuracy, but adding LLM correction reduces this to 78.77\%.

Chunk-based approaches with vision-language models perform poorly. Chunk + MiniCPM-V achieves only 9.36\% CER accuracy with negative accuracy values on several document types, indicating systematic hallucination and over-insertion. Chunk + Qwen3-VL (47.76\% CER) performs better but remains far below conventional OCR baselines.

\subsection{Performance by Document Type}\label{subsec7}

Table~\ref{tab:bytype_all} presents the results for all thirteen pipelines across document types, revealing how production method and visual complexity affect extraction quality.

\begin{table*}[t]
\caption{Average CER and WER accuracy (\%) by document type for all evaluated pipelines.}\label{tab:bytype_all}
\resizebox{\textwidth}{!}{%
\begin{tabular}{lcccccccccc}
\toprule
& \multicolumn{2}{c}{\textbf{Type 1}} & \multicolumn{2}{c}{\textbf{Type 2}} & \multicolumn{2}{c}{\textbf{Type 3}} & \multicolumn{2}{c}{\textbf{Type 4}} & \multicolumn{2}{c}{\textbf{Type 5}} \\
& \multicolumn{2}{c}{\textit{Low Quality}} & \multicolumn{2}{c}{\textit{Multi-Column}} & \multicolumn{2}{c}{\textit{Clean Scan}} & \multicolumn{2}{c}{\textit{Digital}} & \multicolumn{2}{c}{\textit{No Embed}} \\
\cmidrule(lr){2-3}\cmidrule(lr){4-5}\cmidrule(lr){6-7}\cmidrule(lr){8-9}\cmidrule(lr){10-11}
\textbf{Pipeline} & CER & WER & CER & WER & CER & WER & CER & WER & CER & WER \\
\midrule
PyMuPDF                   & 91.40 & 80.12 & 57.27 & 54.49 & 89.98 & 85.51 & 97.52 & 97.60 & 0.00  & 0.00  \\
Docling (No OCR)          & 90.04 & 84.01 & 55.15 & 45.33 & 91.93 & 88.04 & 96.54 & 94.00 & 0.00  & 0.00  \\
Docling + EasyOCR         & 92.90 & 83.91 & 82.39 & 69.51 & 91.97 & 88.07 & 96.54 & 94.00 & 63.13 & 54.85 \\
Docling + Tesseract       & 91.54 & 82.13 & 84.66 & 72.84 & 95.41 & 87.78 & 96.54 & 94.00 & 77.17 & 74.64 \\
Marker                    & 97.79 & 96.26 & 98.13 & 96.62 & 99.56 & 98.00 & 98.32 & 98.05 & 99.70 & 99.59 \\
PyMuPDF + LLM             & 77.39 & 66.51 & 21.36 & 18.34 & 61.36 & 62.74 & 63.70 & 67.67 & 0.00  & 0.00  \\
Docling (No OCR) + LLM    & 86.74 & 81.38 & 50.33 & 46.88 & 83.94 & 81.80 & 87.32 & 83.10 & 0.00  & 0.00  \\
Docling + EasyOCR + LLM   & 86.18 & 79.97 & 78.89 & 72.32 & 72.56 & 66.19 & 85.53 & 81.64 & 49.74 & 36.86 \\
Docling + Tesseract + LLM & 78.57 & 72.49 & 73.42 & 66.72 & 84.82 & 82.29 & 89.05 & 84.23 & 67.97 & 63.80 \\
Marker + LLM              & 90.56 & 87.44 & 90.31 & 87.64 & 51.66 & 45.35 & 89.49 & 84.67 & 92.10 & 91.40 \\
Chunk + MiniCPM-V         & $-$153.60 & $-$155.88 & 32.91 & 23.77 & 36.18 & 26.07 & 70.19 & 63.74 & 61.13 & 52.37 \\
Chunk + Qwen3-VL          & 50.41 & 43.35 & 17.37 & 14.51 & 39.22 & 36.53 & 77.67 & 75.88 & 54.10 & 48.33 \\
Chunk + EasyOCR           & 41.06 & 25.57 & 34.18 & 26.68 & 42.70 & 36.63 & 76.81 & 70.19 & 65.66 & 54.55 \\
\botrule
\end{tabular}%
}
\end{table*}

\textbf{Type 1 (Low-Quality Scans with Annotations):} Marker achieves 97.79\% CER and 96.26\% WER accuracy despite severe degradation and annotations. Conventional OCR approaches perform reasonably (Docling + EasyOCR: 92.90\% CER accuracy; Docling + Tesseract: 91.54\% CER accuracy), and direct extraction methods achieve similar character accuracy (PyMuPDF: 91.40\% CER accuracy) due to the presence of embedded OCR layers in these legacy scans. LLM post-correction degrades results across all base pipelines, and chunk-based approaches perform poorly, with Chunk + MiniCPM-V yielding strongly negative accuracy ($-$153.60\% CER accuracy), indicating massive hallucination and over-insertion of text.

\textbf{Type 2 (Multi-Column Complex Layouts):} Layout complexity substantially degrades performance. Direct approaches fail significantly (PyMuPDF: 57.27\% CER accuracy; Docling No OCR: 55.15\% CER accuracy) due to incorrect reading order. Forcing OCR through Docling improves results considerably (Docling + Tesseract: 84.66\% CER accuracy; Docling + EasyOCR: 82.39\% CER accuracy). Marker maintains robust performance (98.13\% CER accuracy; 96.62\% WER accuracy), demonstrating superior layout understanding. LLM post-correction produces mixed effects in this category: it severely degrades PyMuPDF + LLM (21.36\% CER accuracy), while Docling + EasyOCR + LLM (78.89\% CER accuracy) performs comparably to its base.

\textbf{Type 3 (Clean Scans):} On high-quality single-column scans, most conventional approaches perform well. Marker achieves near-perfect accuracy (99.56\% CER accuracy; 98\% WER accuracy). Conventional OCR reaches strong levels (Docling + Tesseract: 95.41\% CER accuracy; Docling + EasyOCR: 91.97\% CER accuracy), and direct extraction performs adequately when OCR layers are present (Docling No OCR: 91.93\% CER accuracy; PyMuPDF: 89.98\% CER accuracy). However, LLM post-correction again degrades performance substantially in most cases, with PyMuPDF + LLM dropping to 61.36\% CER accuracy.

\textbf{Type 4 (Digital PDFs):} All methods perform well on digital documents with embedded text. Marker achieves 98.32\% CER accuracy and 98.05\% WER accuracy. The other direct extraction approaches reach their best performance (PyMuPDF: 97.52\% CER accuracy; Docling No OCR: 96.54\% CER accuracy), while forced OCR pipelines perform identically to the no-OCR baseline (Docling + Tesseract: 96.54\% CER accuracy; Docling + EasyOCR: 96.54\% CER accuracy), suggesting that the embedded text is of sufficient quality and forcing OCR neither helps nor hurts. LLM post-correction degrades all results substantially, with PyMuPDF + LLM falling to 63.70\% CER accuracy.

\textbf{Type 5 (No Embedded Text):} Scans without embedded text expose fundamental limitations of extraction-based approaches. PyMuPDF and Docling No OCR achieve 0\% accuracy, producing empty outputs. Their LLM-enhanced counterparts also yield 0\%, as there is no text to correct. OCR-based methods show degraded but functional performance (Docling + Tesseract: 77.17\% CER accuracy; Docling + EasyOCR: 63.13\% CER accuracy). Chunk-based approaches achieve moderate results in this category (Chunk + EasyOCR: 65.66\% CER accuracy; Chunk + MiniCPM-V: 61.13\% CER accuracy), suggesting that for documents where no embedded text exists, even these lower-performing approaches provide some value. Marker maintains exceptional accuracy (99.70\% CER accuracy; 99.59\% WER accuracy).

\subsection{Error Pattern Analysis}\label{subsec8}

Qualitative analysis reveals characteristic error patterns for each pipeline family.

\begin{figure}[t]
  \centering
  \begin{subfigure}{\columnwidth}
    \includegraphics[width=\columnwidth]{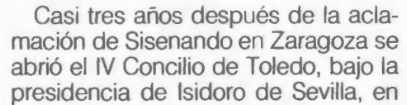}
    \caption{Original text}
  \end{subfigure}
  \begin{subfigure}{\columnwidth}
    \includegraphics[width=\columnwidth]{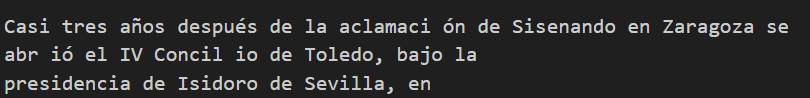}
    \caption{Extracted text}
  \end{subfigure}
  \caption{Word segmentation failures.}
  \label{fig:segmentation}
\end{figure}

\noindent\textbf{Word Segmentation Errors:} Embedded-text extraction and basic OCR frequently introduce spurious spaces within words or merge distinct words. Figure~\ref{fig:segmentation} illustrates ``aclamación'' incorrectly segmented as ``aclama ción'', while words such as ``Concilio'' and ``abrió'' are also erroneously split, introducing noise. These errors severely degrade word-level accuracy despite moderate character-level accuracy.

\begin{figure}[t]
  \centering
  \begin{subfigure}{\columnwidth}
    \includegraphics[width=\columnwidth]{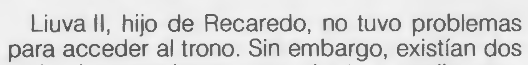}
    \caption{Original text}
  \end{subfigure}
  \begin{subfigure}{\columnwidth}
    \includegraphics[width=\columnwidth]{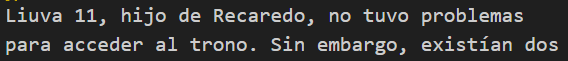}
    \caption{Extracted text}
  \end{subfigure}
  \caption{Roman numeral misrecognition.}
  \label{fig:roman_numeral}
\end{figure}

\noindent\textbf{Character Substitutions:} Visually similar characters are frequently confused: \textit{t}/\textit{l}, \textit{I}/\textit{1}, \textit{O}/\textit{0}, \textit{a}/\textit{u}. Roman numerals prove particularly problematic, with ``II'' consistently misrecognized as ``11''. Additionally, highly distorted character sequences appear, such as ``destruyéndolo'' becoming ``dcslruyéiiidolo'', and symbols like \textit{s} being misrecognized as \textit{\S}. Figure~\ref{fig:roman_numeral} shows the Liuva~II/11 error motivating this study.

\noindent\textbf{Layout Order Corruption:} Multi-column documents exhibit reading order errors where text from adjacent columns is interleaved. Embedded-text approaches particularly struggle, often reading left-to-right across column boundaries rather than top-to-bottom within columns. When the system does not correctly detect the column structure, it may attempt to read the page in a single, linear sequence, ignoring the intended reading order.

\noindent\textbf{Symbol Misinterpretation:} Specialized symbols (\dag, \ddag, \S) are frequently misrecognized as letters. The dagger symbol (\dag) indicating death dates consistently becomes ``t'', eliminating semantic information. Punctuation errors also occur, with commas becoming dots or hyphens.

\noindent\textbf{Visual Noise:} Images, stains, and artifacts generate spurious character sequences. Poor-quality regions can produce strings of random characters as OCR systems attempt to interpret visual noise as text. In image regions, the recognizer may attempt to interpret visual patterns as characters, resulting in the insertion of unexpected symbols into the output.

\noindent\textbf{Footnote Errors:} Footnotes are particularly difficult to extract accurately and often contain the highest concentration of errors. Smaller font sizes make characters harder to distinguish at the pixel level. Superscript numbers preceding footnotes are often mistaken for other numbers, substituted for other symbols, or omitted entirely.

\noindent\textbf{LLM Post-Correction Errors:} Post-correction with LLMs introduces distinct error modes. The model sometimes hallucinates errors absent in the source text, for example, identifying a supposed repetition of a preposition where the actual error is a typographical artifact. Instead of returning corrected text, the model sometimes produces explanations despite explicit instructions. Some paragraphs with heavy OCR corruption prove too complex for the model to correct accurately. Words not well represented in the vocabulary of the model are left uncorrected (e.g., ``vándulos'' remaining instead of ``vándalos''). A significant issue arises with Latin words closely resembling Spanish ones: the model incorrectly normalizes correct Latin forms into Spanish (e.g., ``possessores'' becoming ``poseedores''), introducing errors rather than correcting them.

\noindent\textbf{Vision Model Errors:} Vision models used in chunk-based extraction exhibit additional failure modes. They sometimes add extra text not present in the original block (e.g., prefixing output with labels such as ``TEXTO DE LA IMAGEN:''). Blocks consisting solely of noise are sometimes transcribed as characters rather than identified as non-text. Spelling errors from conventional OCR persist, and the chunking process itself introduces ordering difficulties, particularly for multi-column layouts.

\section{Discussion}\label{sec5}

\subsection{RQ1: Pipeline Performance and Error Patterns}\label{subsec9}

Our results demonstrate clear performance differentiation across the pipeline families analyzed. End-to-end document parsing (Marker) substantially outperforms all alternatives, achieving near-optimal accuracy even on degraded and complex documents. This suggests that joint training of layout detection and text recognition on document-specific corpora yields qualitative improvements over modular pipelines combining generic computer vision and OCR components.

Conventional OCR systems (Tesseract, EasyOCR) show moderate performance with characteristic failure modes: word segmentation errors, character substitutions, and layout corruption on multi-column documents. Performance degrades predictably with scan quality and layout complexity. Embedded-text extraction (PyMuPDF, Docling No OCR) succeeds only when high-quality text layers exist, completely failing on true scans.

Vision-language models show disappointing performance despite theoretical advantages. Their tendency toward hallucination and verbose outputs (including explanatory text rather than pure transcription) makes them unsuitable for production digitization workflows. The Chunk + MiniCPM-V pipeline produced negative accuracy values on Type 1 documents ($-$153.60\% CER), meaning it introduced more erroneous characters than the total length of the reference text. Even the best vision model approach, Chunk + Qwen3-VL (47.76\% CER overall), remains far below conventional OCR baselines. The poor results may reflect misalignment between vision model training objectives (general scene understanding, visual question answering) and the precise character-level accuracy required for OCR.

\subsection{RQ2: Document Heterogeneity}\label{subsec10}

Document stratification reveals systematic performance variation across production methods and visual complexity. Clean digital PDFs (Type 4) and high-quality scans (Type 3) present minimal challenges for most pipelines. Multi-column layouts (Type 2) and degraded scans with annotations (Type 1) expose fundamental limitations of conventional approaches.

The Type 5 results (no embedded text) are particularly instructive. The 0\% accuracy of embedded-text approaches underscores a critical limitation: tools like PyMuPDF and basic Docling configurations silently fail on true scans, producing empty outputs or fragmentary text. This presents serious risks for automated workflows where silent failures may go undetected.

The consistent performance of Marker across all document types ($>$97\% CER accuracy across all categories) suggests that its architectural choices specifically address document heterogeneity. The combination of EfficientViT-based layout detection and Donut-based text recognition, trained end-to-end on diverse documents, appears to generalize effectively across production methods and quality levels.

An interesting observation in the Type 4 (Digital PDF) results is that Docling produces identical results regardless of whether OCR is enabled or not (96.54\% CER for all three Docling configurations). This confirms that for digitally-born documents with clean embedded text, the OCR step is redundant. By contrast, forcing OCR makes a substantial difference on Type 2 (multi-column) and Type 5 (no embedded text) documents, where it can improve or enable extraction entirely.

\subsection{RQ3: LLM Post-Correction}\label{subsec11}

Our results provide evidence that LLM-based post-correction does not systematically improve OCR quality when evaluated using edit-distance metrics. All LLM-augmented pipelines show degraded overall performance relative to their base extraction systems. This finding aligns with \citet{Kanerva2025}, who caution against assumptions of universal LLM benefit for OCR correction. Several factors explain this degradation:

\begin{itemize}
    \item \textbf{Metric Alignment:} Edit-distance metrics penalize any deviation from ground truth, including LLM ``corrections'' of already-accurate text. LLMs trained on modern corpora may normalize archaic spellings, adjust punctuation to contemporary conventions, or ``fix'' correct but unusual phrasings---all counted as errors under strict edit distance.
    \item \textbf{Hallucination:} Despite carefully engineered prompts, LLMs exhibit tendency toward hallucination, inventing errors that do not exist in input text and adding explanatory commentary rather than pure corrections.
    \item \textbf{Non-Determinism:} LLM outputs vary across runs due to sampling, introducing inconsistency unsuitable for reliable digitization pipelines.
    \item \textbf{Vocabulary Coverage:} Historical proper nouns, archaic terms, and Latin phrases are poorly represented in LLM training corpora, leading to incorrect ``normalizations.'' The observed case where correct Latin ``possessores'' was changed to Spanish ``poseedores'' exemplifies this problem.
    \item \textbf{Complex Corruption:} Some OCR-corrupted paragraphs are so severely damaged that the model fails to reconstruct the original meaning, producing only partial corrections while leaving fundamental errors intact.
\end{itemize}

These findings suggest that LLM post-correction may benefit from alternative evaluation frameworks emphasising semantic preservation and readability over strict character-level fidelity. For applications prioritising human readability (e.g., publicly accessible digital files), LLM normalisation might improve user experience despite degrading edit-distance scores. Conversely, for scholarly applications requiring faithful reproduction of original text, LLM correction appears counterproductive.

Our findings align with recent literature on LLM-based OCR correction. Koynov \cite{koynov2025opportunities} documented similar challenges including hallucination and vocabulary coverage gaps, while Kanerva et al. \cite{Kanerva2025} demonstrated that LLM effectiveness varies significantly across languages and base OCR quality levels.

\subsection{Limitations}\label{subsec12}

This study acknowledges some limitations. First, the corpus is limited to Spanish historiographical texts on a specific topic. Generalisation to other languages, domains, and time periods requires validation. Second, manual ground truth construction is labour-intensive, limiting corpus size. Larger-scale evaluation would strengthen findings. Third, we employ edit-distance metrics standard in OCR research but potentially misaligned with downstream application requirements (e.g., semantic search, fact extraction). Fourth, LLM experiments used a single model (Qwen3 8B) with specific prompting strategies; alternative models and prompts might yield different results. Fifth, we do not evaluate commercial OCR services (Google Cloud Vision, AWS Textract, Azure Computer Vision) due to cost and reproducibility constraints, though these may outperform open-source alternatives.

\section{Conclusion}\label{sec6}

This study presents a systematic evaluation of PDF-to-text extraction pipelines for historiographical documents, revealing substantial performance variation across approaches and document types. End-to-end document parsing achieves superior and stable performance (Marker: 98.70\% CER accuracy overall), while conventional OCR degrades on complex layouts and degraded scans. Embedded-text extraction succeeds only on digital documents and silently fails on true scans. LLM-based post-correction does not systematically improve quality under edit-distance evaluation and frequently degrades accurate extractions.

These findings have direct implications for digital humanities practice and AI knowledge systems. As conversational AI and RAG architectures scale to incorporate digitised historical collections, OCR quality emerges as a critical bottleneck. Systematic errors at the digitisation stage propagate through retrieval and generation pipelines, producing outputs that appear confident but contain errors invisible to end users.

Our stratified analysis provides actionable guidance for practitioners. End-to-end parsing proved to be the most reliable approach across all document types and should be the preferred choice when working with heterogeneous historical collections. Since document characteristics such as scan quality, layout complexity, and the presence of embedded text layers have a significant impact on extraction accuracy, workflows should be adapted accordingly rather than applying a single pipeline uniformly. Additionally, LLM-based post-correction should not be assumed to be beneficial by default---our results show that it can introduce new errors, so it should be validated on representative samples before being applied at scale. Improving digitisation quality is not only a technical challenge but also a necessary step to ensure that historical knowledge remains accurate and trustworthy as it becomes increasingly accessed through AI systems.

\backmatter



\section*{Statements and Declarations}

\textbf{Funding:} This work has been supported by the Madrid Government (Comunidad de Madrid-Spain) under the Multiannual Agreement with UC3M (BADDO-CM-UC3M), by the ``Generation of Reliable Synthetic Health Data for Federated Learning in Secure Data Spaces'' Research Project (Agencia Estatal de Investigación (AEI)/European Regional Development Fund (ERDF), EU) funded by Ministerio de Ciencia, Innovación y Universidades (MCIN)/AEI/10.13039/501100011033 under Grant PID2022-141045OB-C43, as well as by the GENIE Learn project under Grant PID2023-146692OB-C31 funded by MICIU/AEI/10.13039/501100011033 and by ERDF/UE.

\end{document}